\documentclass[11pt,twoside]{article}
\usepackage{amssymb,latexsym,amsmath,amstext,amsthm}

\def\aut#1#2{
{\small \noindent
\parbox[t]{6.5cm}{#1}
 \hfill
\parbox[t]{6.5cm}{#2}
} }

\begin{document}

\thispagestyle{plain}

\title{\bf RIGGINGS OF  LOCALLY COMPACT ABELIAN GROUPS.}
\author{M. Gadella, F. G\'omez, S. Wickramasekara.}

\date{}

\maketitle

\begin{abstract}
We obtain a set of generalized eigenvectors that provides a
generalized spectral decomposition for a given unitary
representation of a commutative, locally compact topological group.
These generalized eigenvectors are functionals belonging to the dual
space of a rigging on the space of square integrable functions on
the character group. These riggings are obtained through suitable
spectral measure spaces.
\end{abstract}

\label{first}

\section[]{Introduction}

The purpose of the present paper is to take a first step towards a
general formalism of unitary representations of groups and
semigroups on rigged Hilbert spaces. To begin with, we want to
introduce the theory corresponding to Abelian locally compact
groups, leaving the more general nonabelian case as well as
semigroups for a later work. We recall that a {\it rigged Hilbert
space} or a rigging of a Hilbert space $\cal H$ is a triplet of the
form

\begin{equation}\label{1}
{\bf\Phi}\subset {\cal H}\subset{\bf \Phi}^\times\,,
\end{equation}
where ${\bf\Phi}$ is a locally convex space dense in $\cal H$ with a
topology stronger than that inherited from $\cal H$ and ${\bf
\Phi}^\times$ is the dual space of ${\bf\Phi}$. In this paper, we
shall always assume that $\cal H$ is separable.

To each self adjoint operator $A$ on $\cal H$, the von Neumann
theorem \cite{VN} associates a {\it spectral measure space}. This is
the quadruple $(\Lambda,{\cal A},{\cal H},P)$, where $\cal H$ is the
Hilbert space on which $A$ acts, $\Lambda=\sigma(A)$ is the spectrum
of $A$,  $\cal A$ is the family of Borel sets in $\Lambda$, and  $P$
is the projection valued measure on $\cal A$ determined by $A$
through the von Neumann theorem. Obviously $\Lambda\subset\mathbb
R$. A complete discussion on the relation between these concepts can
be found in \cite{GG}. We say that the topological vector space
$({\bf\Phi},\tau_{\bf\Phi})$ (vector space ${\bf\Phi}$ with the
locally convex topology given by $\tau_{\bf\Phi}$) equips or rigs
the spectral measure $(\Lambda,{\cal A},{\cal H},P)$ if the
following conditions hold:
\begin{enumerate}
\item[{i.}]
There exists a one-to-one linear mapping $I:{\bf\Phi}\longmapsto
{\cal H}$ with range dense in ${\cal H}$. We can assume that
${\bf\Phi}\subset {\cal H}$ is a dense subspace of $\cal H$ and $I$,
the canonical injection from ${\bf\Phi}$ into $\cal H$.

\item[{ii.}]
There exists a $\sigma$-finite measure $\mu$ on $(\Lambda,\cal A)$,
a set $\Lambda_0\subset \Lambda$ with zero $\mu$ measure and a
family of vectors in ${\bf\Phi}^\times$ of the form

\begin{equation}\label{2}
\{|{\lambda k}^\times\rangle\in{\bf\Phi}^\times \,:\, \lambda\in
\Lambda\backslash \Lambda_0,\, k\in\{1,2,\dots,m\}\},
\end{equation}
where $m\in\{\infty,1,2,\ldots\}$, such that

\begin{equation}\label{3}
(\phi, P ( E )\varphi)_{\cal H}=\int_ E  \sum_{k=1}^m
\langle\phi|{\lambda k}^\times\rangle\, \langle\varphi|{\lambda
k}^\times\rangle^\ast\,d\mu(\lambda),\quad
\forall\,\phi,\varphi\in\Phi,\,\,\forall\, E \in\cal A.
\end{equation}
\end{enumerate}
Each family of the form (\ref{2}) satisfying (\ref{3}) is called a
{\it complete system of Dirac kets} of the spectral measure
$(\Lambda,{\cal A},{\cal H},P)$ in $({\bf\Phi},\tau_{\bf\Phi})$. In
this case, the triplet ${\bf\Phi}\subset{\cal H}\subset
{\bf\Phi}^\times$ is a rigged Hilbert space, which is called a {\it
rigging} of $(\Lambda,{\cal A},{\cal H},P)$.

Conversely,  the von Neumann theorem asserts that a projection
valued measure defined on the $\sigma$-algebra of Borel sets on a
subset of the real line determines a self adjoint operator $A$. If
$(\Lambda,{\cal A},{\cal H},P)$, where $\Lambda\subset{\mathbb{R}}$,
is such a measure space, then for $\varphi$ and $\phi$ on a suitable
dense domain, the self-adjoint operator $A$ such that
$\Lambda=\sigma(A)$ is defined by
\begin{equation}
(\phi,A\varphi)_{\cal H}=\int_{\Lambda} \sum_{k=1}^m \lambda\,
\langle\phi|{\lambda k}^\times\rangle\, \langle\varphi|{\lambda
k}^\times\rangle^\ast\,d\mu(\lambda)\
\end{equation}
where $\mu$, $|\lambda k^\times\rangle$ and $m$ are as defined in
\eqref{3}. Further, if $f(\lambda)$ is a measurable  complex valued
function on $\Lambda$, then, for $\phi,\varphi$ on a suitable dense
domain, which is the whole of $\cal H$ if $f(\lambda)$ is bounded,
the operator valued function $f(A)$ is defined by
\begin{equation}\label{4}
(\phi,f(A)\varphi)_{\cal H}=\int_{\Lambda} \sum_{k=1}^m f(\lambda)\,
\langle\phi|{\lambda k}^\times\rangle\, \langle\varphi|{\lambda
k}^\times\rangle^\ast\,d\mu(\lambda)\,.
\end{equation}
The functionals $|{\lambda k}^\times\rangle\in{\bf\Phi}^\times$ and
the complex numbers $f(\lambda)$ are the generalized eigenvectors
and respective generalized eigenvalues of $f(A)$ \cite{BG}. In
particular, if $f(\lambda)=e^{it\lambda}$, where $t\in\mathbb R$,
the set of operators $e^{itA}$ forms a one parameter commutative
group of unitary operators and ${\bf\Phi}\subset{\cal H}\subset
{\bf\Phi}^\times$ as defined above is a rigging for this group.

One can expect that  similar riggings exist for unitary
representations of arbitrary groups and semigroups and that the
operators of the representations can be expanded in terms of
generalized eigenvectors and eigenvalues as in (\ref{4}). Riggings
that make use of Hardy functions on a half plane exist for one
parameter dynamical semigroups  $e^{-itH},\ t\le 0$ and  $e^{-itH},\
t\ge 0$, where $H$ is the Hamiltonian \cite{BG}.

In the present paper, we show that riggings along the above lines
always exist for unitary representations of Abelian  locally compact
groups. In particular, let $G$ be an Abelian locally compact group
and $\pi$, a unitary representation of $G$ on a separable Hilbert
space $\cal H$. We will see
 that the Fourier transform on $G$, or equivalently, the
Gelfand transformation on the $C^*$-algebra $L^1(G)$ allows us to
represent $\pi$ in terms of {\it generalized eigenfunctions} and
riggings of $\cal H$ in a manner similar to the description given in
\cite{GG} for the action of a spectral measure.

\section{Characters of Abelian Locally Compact Groups.}

Let $G$ be a locally compact abelian group with Haar measure $\mu$.
A character $\chi$ of $G$ is any continuous mapping from $G$ into
the set of complex numbers $\mathbb C$ such that
$\chi(g_1g_2)=\chi(g_1)\chi(g_2)$ for all $g_1,g_2\in G$ and
$|\chi(g)|=1$ for all $g\in G$, i.e., a character of $G$ is a
continuous homomorphism from $G$ into the unit circle $\mathbb T$.
The set of all the characters of $G$ forms a group, $\widehat G$,
which is often called the dual group of $G$. We shall use the
notation $\chi(g):=\langle g|\chi\rangle$.

Let $L^1(G)$ be the space of complex valued functions, integrable in
the modulus with respect to the Haar measure $\mu$ on $G$. $L^1(G)$
is an abelian $*$-algebra, with the convolution product. The dual
group $\widehat G$ can be identified with the set of maximal ideals
of $L^1(G)$ \cite{N}. When endowed with the Gelfand topology,
$\widehat G$ is a compact Hausdorf space (see \cite{R} page 268).

For any $\chi\in\widehat{G}$, we may define a linear functional
$\Lambda_\chi$ on $L^1(G)$ by
\begin{equation}
\Lambda_\chi(f)=\int\langle g|\chi\rangle^*f(g)d\mu(g)\,.\label{4.5}
\end{equation}

Let $C(\widehat G)$ be the space of complex continuous functions on
$\widehat G$ with the supremun norm topology. The Gelfand-Fourier
transform is the mapping ${\cal F}: L^1(G)\longmapsto C(\widehat G)$
defined by:
\begin{equation}\label{5}
    [{\cal F}f](\chi)=\widehat f(\chi)=\Lambda_\chi(f)=\int \langle
    g|\chi\rangle^*\,f(g)\,d\mu(g)\,.
\end{equation}

Let $(\pi,{\cal H})$ be a unitary representation of $G$. Then (see
\cite{F} page 105), there is a unique spectral measure $(\widehat
G,{\cal B},{\cal H},P)$, where $\cal B$ is the $\sigma$-algebra of
Borel sets on $\widehat G$, such that for all $g\in G$ and all $f\in
L^1(G)$, we have

\begin{equation}\label{6}
\pi(g)=\int_{\widehat G}\langle g|\chi\rangle\,dP(\chi)\qquad;\qquad
\pi(f)=\int_{\widehat G} \Lambda_\chi(f)\,dP(\chi)\,.
\end{equation}
There is a one to one correspondence between unitary representations
of $G$ and non degenerate $*$-representations\footnote{Non
degenerate means that $\pi(f)v={\bf 0}$ for every $f$ implies $v=\bf
0$. The representation has also the property that
$\pi(f^*)=\pi^\dagger(f)$, where $f\mapsto f^*$ is the involution on
$L^1(G)$, see \cite{F}.} of $L^1(G)$ as given by (\ref{5}) and
(\ref{6}).

\subsection{Riggings of functions of characters.}

Let us consider the spectral measure space $(\widehat G,{\cal
B},{\cal H},P)$ introduced in the previous section. For simplicity
in the discussion, we assume the existence of a cyclic vector
$u\in\cal H$. This means that the subspace spanned by the vectors of
the form $P(E)u$ with $E\in\cal A$ is dense in $\cal H$. The general
case can be easily obtained as a finite or countable direct sum of
cyclic subspaces of $\cal H$.

Then, the von Neumann decomposition theorem \cite{VN} establishes
that being given the spectral measure space $(\widehat G,{\cal
B},{\cal H},P)$ and a positive measure $\nu$ on $(\widehat G,{\cal
B})$ with maximal spectral type\footnote{For a definition and
properties of the spectral type, see \cite{VN,GG}.} $[P]$ ($\nu\in
[P]$), there exists a unitary mapping $U:{\cal H}\longmapsto
L^2(\widehat G,d\mu)\,,$ such that $\pi_\nu(g):=U\pi(g)U^{-1}$ is
the multiplication by $\langle g|\chi\rangle$ on $L^2(\widehat
G,d\nu)$:
\begin{equation}\label{8}
\pi_\nu(g) \phi(\chi)=U\pi(g)U^{-1} \phi(\chi)= \langle
g|\chi\rangle \,\phi(\chi)\,,\qquad \forall\, \phi(\chi)\in
L^2(\widehat G,d\nu)\,.
\end{equation}
Since $\pi_\nu(g)$ is a multiplication operator, it is easy to see
that the Dirac delta type Radon measures $\lambda(\chi)\delta_\chi$
form a complete system of Dirac kets for the spectral measure space
$(\widehat G,{\cal B},{\cal H},P)$ in the sense given by (\ref{3}).
For any $f(\chi)\in\bf\Phi$ these deltas satisfy
\begin{equation}\label{9}
\int_{\widehat G} f(\chi)\,\delta_{\chi'}\,d\nu=f(\chi')\,.
\end{equation}
Thus, a possible choice for $\bf\Phi$ is $C(\widehat G)$, the space
of continuous functions on $\widehat G$ endowed with a topology
$\tau_{\bf\Phi}$ stronger than both the topologies of the supremun
and the $||\cdot||_{L^2(\widehat G,d\nu)}$ norm. In this case, the
dual $\bf\Phi^\times$ of $\bf\Phi$ includes the space of all Radon
measures on $(\widehat G,{\cal B})$. We have the rigged Hilbert
space ${\bf\Phi}\subset L^2(\widehat G)\subset \bf\Phi^\times$.

\section{Positive Type Functions and Riggings.}

Next, we shall introduce another representation $\pi_\phi$ of $G$
linked to a function of  positive type, that can be defined as
follows: Let $\phi(g)\in L^\infty (G)$. We say that $\phi(g)$ is a
function of {\it positive type} if for any $f(g)\in L^1(G)$, we have
that
\begin{equation}\label{10}
\int_G\int_G f^*(g)f(gg')\phi(g')\,d\mu(g)\,d\mu(g')\ge 0\,,
\end{equation}
where the star $*$ denotes complex conjugation.

If $\phi(g)$ is a function of positive type, then, the following
positive Hermitian form on $L^1(G)$
\begin{equation}\label{11}
    \langle h|f\rangle_\phi:= \int_G\int_G
    h^*(g')f(g)\phi(g^{-1}g')\,d\mu(g')\,d\mu(g)
\end{equation}
is semi-definite in the sense that it may exist non-zero functions
$f\in L^1(G)$ such that $\langle f|f\rangle_\phi=0$. These functions
form a subspace of $L^1(G)$ that we denote by $\cal N$. Consider the
factor space $L^1(G)/{\cal N}$ and again denote by $\langle
\cdot|\cdot\rangle_\phi$ the scalar product induced on $L^1(G)/{\cal
N}$ by the Hermitian form (\ref{11}). The completion of
$L^1(G)/{\cal N}$ by $\langle \cdot|\cdot\rangle_\phi$ gives a
Hilbert space usually denoted as ${\cal H}_\phi$. Then, for any
$g\in G$ and $f(g)\in L^1(G)$, we define:
\begin{equation}\label{12}
    (L_g)f(g'):= f(g^{-1}g')\,.
\end{equation}
Note that $L_g$ preserves the scalar product $\langle
\cdot|\cdot\rangle_\phi$:
\begin{eqnarray}
\langle L_g h|L_g f\rangle_\phi= \int_G\int_G
h^*(g^{-1}g')f(g^{-1}g'')\phi(g^{'-1}g'')\,d\mu(g')\,d\mu(g'')\nonumber\\[2ex]
= \int_G\int_G h^*(g') f(g'')\phi((g g')^{-1}(g
g''))\,d\mu(g')\,d\mu(g'')=\langle h|f\rangle_\phi\,, \label{13}
\end{eqnarray}
for all $f(g)\in L^1(G)$. This also shows that $L_g{\cal
N}\subset{\cal N}$ and therefore $L_g$ induces a transformation on
the factor space $L^1(G)/{\cal N}$, that we also denote as $L_g$,
defined as
\begin{equation}\label{14}
L_g(f(g')+{\cal N}):= f(g^{-1}g')+{\cal N}=L_g(f(g'))+{\cal N}\,.
\end{equation}
By (\ref{13}), we easily see that $L_g$ preserves the scalar product
on $L^1(G)/{\cal N}$. It is obviously invertible. Therefore, it can
be uniquely extended into a unitary operator on ${\cal H}_\phi$.
Then, if for each $g\in G$ we write
\begin{equation}\label{15}
\pi_\phi(g)f:=L_g f\,,\qquad \forall\,f\in {\cal H}_\phi\,,
\end{equation}
then, $\pi_\phi(g)$ determines a unitary representation of $G$ on
${\cal H}_\phi$. The proof of this statement is straightforward.

The representation $\pi_\phi(g)$ of $G$ on ${\cal H}_\phi$ can be
lifted to a unitary representation of the group algebra $L^1(G)$ on
${\cal H}_\phi$ that we shall also denote as $\pi_\phi$. In this
case, for all $f\in L^1(G)$, we have $\pi_\phi(f)h:= f*h$. Here, $*$
denotes convolution.

The existence of a cyclic vector $\eta\in{\cal H}_\phi$ for the
representation $\pi_\phi$ is proven in \cite{F}. Recall that $\eta$
is cyclic vector if
 the subspace $\{ \pi_\phi(f)\eta,\ \exists f\in L^1(G)\}$ is dense in
${\cal H}_\phi$. In addition, this result also gives the following
formula that allows to find the function $\phi(g)$ in terms of
$\eta$ and the unitary representation $\pi_\phi$ of $G$ on ${\cal
H}_\phi$:
\begin{equation}\label{17}
\phi(g)=\langle \eta|\pi_\phi(g)\eta\rangle\,.
\end{equation}

Now, let us consider the unitary $\pi_\nu$ representation of $G$
given by (\ref{8}) with cyclic vector $\xi$ and define the following
complex valued function on $G$:
\begin{equation}\label{18}
    \phi(g^{-1}g'):=
\langle\xi|\pi_\nu(g^{-1}g')\xi\rangle_{L^2(\widehat G,d\nu)}
=\langle
    \pi_\nu(g)\xi|\pi_\nu(g')\xi\rangle_{L^2(\widehat G,d\nu)}\,.
\end{equation}
Then, as shown in \cite{F}, Chapter 3, \\
i.) the function $\phi$ is of positive type in the sense of
(\ref{10}), and\\
ii.) the representation of $G$ on ${\cal H}_\phi$ given by
$\pi_\phi$, where $\phi$ is as (\ref{18}) is equivalent to
$\pi_\nu$.

 Note that this result
implies in particular that for this $\phi$ as in (\ref{18})
\begin{equation}\label{19}
\phi(g)=\langle \eta|\pi_\phi(g)\eta\rangle_\phi= \langle
\xi|\pi_\nu(g)\xi\rangle_{L^2(\widehat
G,d\nu)}\,,\qquad\forall\,g\in G\,.
\end{equation}
According to (\ref{8}) and (\ref{18}), we have that
\begin{equation}\label{20}
\phi(g^{-1}g')=\langle \pi_\nu(g)
\xi|\pi_\nu(g')\xi\rangle_{L^2(\widehat G,d\nu)}= \int_{\widehat G}
[\langle g|\chi\rangle\,\xi(\chi)]^*\, \langle g'|\chi\rangle
\,\xi(\chi)\,d\nu(\chi)\,.
\end{equation}
If we carry this formula into (\ref{11}) and apply the Fubini
theorem of the change of the order of integration, we have for all
$f,h\in L^1(G)$:
\begin{eqnarray}
\langle f|h\rangle_\phi=  \int_{\widehat G}\left(\int_G[f(g)\langle
g|\chi\rangle]^*\,d\mu(g)\right)\left(\int_G h(g')\langle
g'|\chi\rangle\,d\mu(g')\right)\,|\xi(\chi)|^2\,d\nu(\chi) \nonumber
\\[2ex]
=\int_{\widehat G} [\widehat f(\chi)]^*\,\widehat
h(\chi)\,|\xi(\chi)|^2\,d\nu(\chi)=\int_{\widehat G} \langle
\widehat f|\chi\rangle\langle \chi|\widehat h\rangle\,
|\xi(\chi)|^2\,d\nu(\chi)\,.\qquad \label{21}
\end{eqnarray}
This latter formula shows that the generalized eigenvalues $F_\chi$
of $\pi_\phi(g)$ are the following: if $f\in {\bf\Phi}:= L^1(G)\cap
L^2(G)$

\begin{equation}\label{22}
|F_\chi\rangle\equiv F_\chi: f\longmapsto |\eta(\chi)|\widehat
f^*(\chi)= |\eta(\chi)| \int_G \langle g|\chi\rangle
f^*(g)\,d\mu(g)\,.
\end{equation}
We endow $\bf\Phi$ with any topology stronger than the topologies
$L^1(G)$ and $L^2(G)$. For instance, we can choose a locally convex
topology with the seminorms $p_1(f):=||f||_{L^1(G)}$ and
$p_2(f):=||f||_{L^2(G)}$, for all $f\in\bf\Phi$.  With this topology
or another stronger one, the antilinear functional $F_\chi$ is
continuous. Then, if we use (\ref{6}) in the scalar product on
${\cal H}_\phi$, we have:
\begin{eqnarray}
 \langle  f|\pi_\phi(g) h\rangle_\phi =  \int_{\widehat G}\langle
g|\chi\rangle
 \,d\langle f|P(\chi)h\rangle_\phi =  \int_{\widehat G} \langle
g|\chi\rangle
 \, |\eta(\chi)|^2 \widehat f^*(\chi)\widehat g(\chi)\,d\nu(\chi)
\nonumber \\[2ex]
= \int_{\widehat G} \langle g|\chi\rangle \, \langle f|F_\chi\rangle
\langle F_\chi|h\rangle\,d\nu(\chi)\,.\qquad \label{23}
\end{eqnarray}
If we omit the arbitrary $f,h\in \bf\Phi$ in (\ref{23}), we have the
following spectral decomposition for $\pi_\phi(g)$ for all $g\in G$:
\begin{equation}\label{24}
\pi_\phi(g)=\int_{\widehat G}
\langle\chi|g\rangle\,|F_\chi\rangle\langle F_\chi|\,d\nu(\chi)\,.
\end{equation}
Note that in the antidual space ${\bf\Phi}^\times$, the generalized
eigenvalue equation $\pi_\phi(g)|F_\chi\rangle= \langle\chi|g\rangle
|F_\chi\rangle$ is valid, where we use the same notation
$\pi_\phi(g)$ for the extensions of these unitary operators into
${\bf\Phi}^\times$.

In conclusion, for each unitary representation of a locally compact
Abelian topological group, we have found an equivalent
representation and a rigged Hilbert space such that each of the
unitary operators of the representation admits a generalized
spectral decomposition in terms of generalized eigenvectors of them.
The eigenvectors of the decomposition are labeled by the group
characters only and their respective eigenvalues, complex numbers
with modulus one, depend on both the corresponding character and the
group element. The spectral decomposition and the corresponding
rigging comes after the existence of a spectral measure space.

Note that the Abelian property is crucial in our derivation and in
particular in the existence of the spectral measure space $(\widehat
G,{\cal B},{\cal H},P)$, since then, the group algebra is also
Abelian and the Gelfand theory applies. An extension of the present
formalism to nonabelian locally compact groups will require an
extension of the Gelfand formalism that at least allows for a new
and consistent definition of the Gelfand Fourier transform
(\ref{5}), an essential feature of our construction.

\section*{Acknowledgements}
We acknowledge the financial support from the Junta de Castilla y
Le\'on Project VA013C05 and the Ministry of Education and Science of
Spain, projects MTM2005-09183 and FIS2005-03988. S.W.~acknowledges
the financial support from the University of Valladolid where he was
a visitor while this work was done and additional financial support
from Grinnell College.

\aut{
M. Gadella \\
Departamento de  F\'isica Te\'orica\\
Facultad de Ciencias\\
c. Real de Burgos, s.n.\\
47011 Valladolid, Spain\\
{\it E-mail address}:\\
 {\tt manuelgadella@yahoo.com.ar}}
{F. G\'omez \\
 Departamento de An\'alisis Matem\'atico \\
 Facultad de Ciencias\\
c. Real de Burgos, s.n.\\
47011 Valladolid, Spain\\
{\it E-mail address}:\\
{\tt fgcubill@am.uva.es}}

\smallskip\noindent
{S. Wickramasekara\\
Department of Physics,\\ Grinnell College, Grinnell, IA 50112, USA\\
{\it E-mail address}:\\
{\tt WICKRAMA@Grinnell.EDU}}

\label{last}


\begin{thebibliography}{99}\itemsep=-.2pc


\bibitem{VN} J. von Neumann, {\it Mathematical Foundations of Quantum
 Mechanics} (Princeton University, Princeton, N.J., 1955).

\bibitem{GG} M. Gadella, F. G\'omez, {\it Foundations of Physics}, {\bf
32}, 815 (2002); M. Gadella, F. G\'omez, {\it International Journal
of Theoretical Physiscs}, {\bf 42}, 2225-2254 (2003); M. Gadella, F
G\'omez, {\it Bulletin des Sciences Math\`ematiques}, {\bf 129}, 567
(2005); M. Gadella, F. G\'omez, {\it Reports on Mathematical
Physics}, {\bf 59}, 127 (2007).

\bibitem{BG}  A. Bohm, M. Gadella, {\it Dirac kets, Gamow vectors  and
Gelfand triplets},  {\it Springer Lecture Notes in Physics}, {\bf
348} (Springer, Berlin 1989).

\bibitem{N} M.A. Naimark, {\it Normed Rings} (Wolters-Noordhoff,
Groningen, The Netherlands, 1970).

\bibitem{R} W. Rudin, {\it Functional Analysis} (McGraw-Hill, New
York 1973).

\bibitem{F} G. B. Folland, {\it A Course in Abstract Harmonic
Analysis} (CRC, Boca Raton, London, 1995).



\end{thebibliography}
\end{document}